\documentclass[twocolumn,showkeys,showpacs,preprintnumbers,prd,superscriptaddress,nofootinbib]{revtex4-2}

\usepackage{graphicx}
\usepackage{epsf}
\usepackage{bm}
\usepackage{amsmath}
\usepackage{amsfonts}
\usepackage{amssymb}
\usepackage{epstopdf}
\usepackage{natbib}
\usepackage{hyperref}
\usepackage{color}
\usepackage{verbatim}
\usepackage{multirow}
\usepackage{bm}
\usepackage{hyperref}

\def\Mpl{M_{\rm P}}

\begin{document}
\preprint{YITP-21-55}

\title{Minimal theory of massive gravity in the light of CMB data and the $S_8$ tension}

\author{Jos\'e C.\ N.\ de Araujo}
\email{jcarlos.dearaujo@inpe.br}
\affiliation{Divis\~ao de Astrof\'isica, Instituto Nacional de Pesquisas Espaciais, Avenida dos Astronautas 1758, S\~ao Jos\'e dos Campos, 12227-010, SP, Brazil}

\author{Antonio De Felice}
\email{antonio.defelice@yukawa.kyoto-u.ac.jp}
\affiliation{Center for Gravitational Physics, Yukawa Institute for Theoretical Physics, Kyoto University, 606-8502, Kyoto, Japan}

\author{Suresh Kumar}
\email{suresh.math@igu.ac.in}
\affiliation{Department of Mathematics, Indira Gandhi University, Meerpur, Haryana-122502, India}
\affiliation{Department of Mathematics, BITS Pilani, Pilani Campus, Rajasthan-333031, India}

\author{Rafael C.\ Nunes}
\email{rafadcnunes@gmail.com}
\affiliation{Divis\~ao de Astrof\'isica, Instituto Nacional de Pesquisas Espaciais, Avenida dos Astronautas 1758, S\~ao Jos\'e dos Campos, 12227-010, SP, Brazil}

\begin{abstract}
We investigate the Minimal Theory of Massive Gravity (MTMG) in the light of different observational data sets which are in tension within the $\Lambda$CDM  cosmology.  In particular, we analyze MTMG model, for the first time, with the Planck-CMB data, and how these precise measurements affect the free parameters of the theory. The MTMG model can affect the CMB power spectrum at large angular scales and cause a suppression on the amplitude of the matter power spectrum. We find that on adding Planck-CMB data, the graviton has a small, positive, but non-zero mass at 68\% confidence level, and from this perspective, we show that the tension between redshift space distortions measurements and Planck-CMB data in the parametric space $S_8 - \Omega_m$ can be resolved within the MTMG scenario. Through a robust and accurate analysis, we find that the $H_0$ tension between the CMB and the local distance ladder measurements  still remains but can be reduced to $\sim3.5\sigma$ within the MTMG theory. The MTMG is very well consistent with the CMB observations, and undoubtedly, it can serve as a viable candidate amongst other modified gravity theories.
\end{abstract}



\maketitle

\section{Introduction}
Over the past two decades, a large volume of cosmological information has been obtained by several data surveys, making the estimates of cosmological parameters increasingly accurate. The understanding of the cosmological probes has been well modeled through the standard $\Lambda$-Cold Dark Matter scenario (the $\Lambda$CDM scenario). The $\Lambda$CDM cosmological model provides a wonderful fit to the current cosmological data, but recently a few tensions and anomalies became statistically significant while analyzing different data sets.  The most discussed tension in the literature is in the estimation of the Hubble constant, $H_0$, between the CMB and the direct local distance ladder measurements. Assuming the $\Lambda$CDM scenario, Planck-CMB data analysis provides $H_0=67.4 \pm 0.5$ km s$^{-1}$Mpc$^{-1}$ \cite{2020}, which is in $4.4\sigma$ tension with a cosmological model-independent local measurement $H_0 = 74.03 \pm 1.42$ km s$^{-1}$Mpc$^{-1}$ \cite{Riess_2019} from the Hubble Space Telescope (HST) observations of 70 long-period Cepheids in the Large Magellanic Cloud. Additionally, a combination of time-delay cosmography from H0LiCOW lenses and the distance ladder measurements are in $5.2\sigma$ tension with the Planck-CMB constraints \cite{Wong_2019}. Recently, based on a joint analysis from several geometrical probes, a tension of $\sim 6\sigma$ was observed in \cite{Bonilla_2021} under the consideration of minimal theoretical assumptions. Motivated by these observational discrepancies, it has been widely discussed in the literature whether a new physics beyond the standard cosmological model can solve the $H_0$ tension (see \cite{Di_Valentino_2021H0,divalentino2021realm, perivolaropoulos2021challenges} and reference therein for a review).
In addition to the $H_0$ disagreement, a significant tension between Planck data with redshift surveys data  and weak lensing measurements has been reported, about the value of the matter density $\Omega_m$, and the amplitude of growth of the structures, $\sigma_8$, also quantified in terms of the parameter: $S_8 = \sigma_8 \sqrt{\Omega_m/0.30}$. A higher $S_8$ value is estimated from CMB data assuming $\Lambda$CDM, namely, $S_8 = 0.834 \pm 0.016$ \cite{2020} from Planck data and $S_8 = 0.840 \pm 0.030$ from ACT+WMAP joint analysis \cite{Aiola_2020}. The cosmic shear and redshift space distortions (RSD) measurements have predicted a lower value of $S_8$. This tension is above the 2$\sigma$ level with KiDS-450 ($S_8 = 0.745 \pm 0.039$) \cite{Joudaki_2017}, KiDS+VIKING-450 ($S_8 = 0.737^{+0.040}_{-0.036}$) \cite{Joudaki_2017_2} and DES ($S_8 = 0.783 \pm 0.021)$ \cite{Troxel_2018}. The KiDS-1000 team reported a 3$\sigma$ tension with Planck-CMB \cite{heymans2020kids1000}. The tension becomes 3.2$\sigma$ if we consider the combination of VIKING-450 and DESY1 \cite{Asgari_2020} and 3.4$\sigma$ for BOSS+VIKING-450 ($S_8 = 0.728 \pm 0.026)$ \cite{Tr_ster_2020}. Also, in agreement with a lower value, there is an estimate from the BOSS Galaxy Power Spectrum $S_8 = 0.703 \pm 0.045$ \cite{Ivanov_2020}. See \cite{Di_Valentino_2021} and references therein for a mini-review and additional information on $S_8$ estimations. Although this tension could be due to systematic errors, it is worthwhile to investigate the possibility of new physics beyond the standard model to explain the $S_8$ tension \cite{Kumar_2019a,Kumar_2019b,Lambiase_2019,Barros_2020,abellan2020hints,Heimersheim_2020,Choi_2021,Skara_2020,marra2021rapid,kumar21,lucca2021dark,Yang_2021,Yang_2020,Choi_2021,choi2021projecting, Kazantzidis_2018}. Additionally, the authors in \cite{Skara_2020} identified a large tension between RSD and CMB measurements. Disagreement between CMB and combinations of RSD measurements with other datasets, including the $E_G$ statistic, is discussed in \cite{nunes2021arbitrating}, pointing out a tension up to 5$\sigma$, depending on the datasets. 

On the other hand, there are theoretical and observational reasons to believe that general relativity (GR) should be modified when gravitational fields are strong and/or on large scales. From an observational point of view, the physical mechanism responsible for accelerating the expansion of Universe at late times is still an open question, and it has been intensively investigated whether modified gravity scenarios can explain such an accelerated stage, as well as to fit very well with the observational data from different sources (see \cite{Ishak_2018} for review). Also, theories beyond GR can serve as alternatives to explain the current $H_0$ tension \cite{DeFelice:2020cpt, Ballardini_2020, Nunes_2018, D_Agostino_2020, Felice_2020, Alestas_2020}. One of the most interesting possibilities for modification of gravity is to give a mass to the graviton (see \cite{de_Rham_2014} for review about massive gravity theories). In this work, we investigate in detail the observational feasibility of the \texttt{Minimal Theory of Massive Gravity} (MTMG) \cite{DeFelice_2015,Felice_2016}, due to its infrared Lorentz violations measurable at cosmological scales. We robustly constrain the MTMG framework using observational probes of the $f\sigma_8$ growth rate of cosmological perturbations, which is a useful bias free statistical cosmic test, as well as we combine these growth rate data with several geometric observations such as the most recent data from Type Ia supernovae and baryon acoustic oscillations.  We derive and discuss observational perspectives from how the MTMG scenario predict the statistical plane for $S_8 - \Omega_m$ in direct comparison with the $\Lambda$CDM model, as well as the potential of the model in alleviating the $H_0$  tension. Furthermore, we investigate for the first time how the Planck-CMB data constrains the MTMG baseline model. Our results also provide new constraint on the graviton mass from Planck-CMB data, obtained in the MTMG context. We find that the use of Planck-CMB data selects parameter space for MTMG which is compatible with a non-zero but positive value of $\mu^2/H_0^2=0.25^{{+}0.16}_{{-}0.10}$, $\mu$ being the mass of the graviton. This is probably the newest and most interesting result from this study.

This paper is structured as follows. In Section \ref{model}, we review and introduce the MTMG scenario. In Section \ref{data}, we present the data sets and methodology used in this work. In section \ref{results}, we discuss the main results of our analysis. In section \ref{final}, we summarize main findings of this study.

\section{Minimal Theory of Massive Gravity}
\label{model}
The model we study here is a model which has been introduced in order to fix the cosmology of massive gravity, i.e., to have a stable FLRW background without ghosts or strong-coupling issues. The model has been introduced in order to study a standard cosmological phenomenology to the theories with a massive graviton. In order for this goal to be achieved, the theory in its simplest form (i.e., in vacuum), has only two gravitational degrees of freedom (corresponding to the two polarization gravitational waves). Furthermore, on FLRW, MTMG shares the same background of dRGT \cite{de_Rham_2010,de_Rham_2011}. This fact was implemented by construction in MTMG. Therefore also MTMG is endowed, as dRGT, with two branches which are named as 1) the self-accelerated branch; and 2) the normal branch.

These two branches have, in general, different FLRW background and perturbation dynamics. In particular, the self-accelerated branch has a phenomenology which is identical in everything to the standard $\Lambda$CDM except for the propagation of tensor modes which are now massive (compared to GR). On the other hand, in the normal branch we have the possibility of giving a general dynamics for the graviton mass. This allows for many phenomenological possibilities. However, in order to study this branch, it is somehow convenient to study its simplest realization, namely the one for which the effective energy density of the MTMG component becomes a constant during the whole evolution.

More in detail, MTMG relies on the presence and choice of a three-dimensional fiducial metric which, in unitary gauge, will be written here as ${\tilde\gamma}_{ij}=\tilde{a}^2\,\delta_{ij}$, where $\tilde{a}=\tilde{a}(t)$ is the fiducial scale factor. On top of that, there is need of the fiducial lapse, $M=M(t)$ and of another 3D tensor $\tilde{\zeta}^i{}_j$, which represents the time variation of the fiducial vielbein (for details, see \cite{DeFelice_2015,Felice_2016}). Out of the fiducial and physical metrics, one builds up the following quantities:
\begin{eqnarray}
\mathcal{K}^k{}_n &=& (\sqrt{{\tilde\gamma}^{-1} \gamma})^k{}_n\,,\\
\mathfrak{K}^k{}_n&=& (\sqrt{{\gamma}^{-1} {\tilde\gamma}})^k{}_n\,,\\
\mathfrak{K}^k{}_n \mathcal{K}^n{}_m&=&\delta^k{}_m=\mathcal{K}^k{}_n \mathfrak{K}^n{}_m\,.
\end{eqnarray}
Out of these quantities, one defines the following symmetric tensor
\begin{eqnarray}
    \Theta^{ij} &=& \frac{\sqrt{\tilde\gamma}}{\sqrt{\gamma}}\{
    c_1 (\gamma^{il}\mathcal{K}^j{}_l+\gamma^{jl}\mathcal{K}^i{}_l)\nonumber\\
    &+& c_2 [\mathcal{K}(\gamma^{il} \mathcal{K}^j{}_l+\gamma^{jl}\mathcal{K}^i{}_l)
    -2{\tilde\gamma}^{ij}]\}+2c_3\gamma^{ij}\,,
\end{eqnarray}
where $c_{1,2,3}$ are constants. We then need the following building blocks
\begin{eqnarray}
\mathcal{C}_0&=&\frac{1}{2}\,m^2\,M\,K_{ij} \Theta^{ij} \nonumber\\
&-&m^2 M \left\{ \frac{\sqrt{\tilde\gamma}}{\sqrt{\gamma}}[c_1\tilde\zeta
+c_2(\mathcal{K}\zeta-\mathcal{K}^m{}_n{\tilde\zeta}^n{}_m)]\right.\nonumber\\
&+&\left. c_3\mathfrak{K}^m{}_n{\tilde\zeta}^n{}_m \right\},\\
\mathcal{C}^n{}_i&=&-m^2 M \left\{ 
\frac{\sqrt{\tilde\gamma}}{\sqrt{\gamma}}
\bigl[\tfrac12(c_1+c_2\mathcal{K})(\mathcal{K}^n{}_i+\gamma^{nm}\mathcal{K}^l{}_m\gamma_{li})\right.\nonumber\\
&-&\left. c_2{\tilde\gamma}^{nl}\gamma_{li}\bigr]+c_3\delta^n{}_i\right\},
\end{eqnarray}
where $\gamma$ and $\tilde\gamma$ represent the determinants of $\gamma_{ij}$ and ${\tilde\gamma}_{ij}$, respectively. Furthermore $K_{ij}$ is the extrinsic curvature of $\gamma_{ij}$, namely $K_{ij}=\frac1{2N}({\dot\gamma}_{ij}-\mathcal{D}_iN_j-\mathcal{D}_jN_i)$, where $N$ is the lapse, $N^i$ the shift vector and $\mathcal{D}$ is the covariant derivative compatible with the 3D-metric $\gamma_{ij}$. Finally, $\mathcal{K}\equiv\mathcal{K}^n{}_n$ and ${\tilde\zeta}\equiv{\tilde\zeta}^n{}_n$. We are now ready to define the action of MTMG as follows
\begin{eqnarray}
    S&=&S_{\rm pre}+\frac{\Mpl^2}{2}\int d^4x N \sqrt\gamma \left(\frac{m^2}{4}\frac{M}{N}\,\lambda\right)^{\!2} (\Theta_{ij}\Theta^{ij}-\tfrac12\Theta^2)\nonumber\\
    &-&\frac{\Mpl^2}{2}\int d^4x \sqrt{\gamma} [\lambda \mathcal{C}_0 - (\mathcal{D}_n \lambda^i) \mathcal{C}^n{}_i] + S_{\rm mat}\,,
\end{eqnarray}
where $S_{\rm mat}$ represents the standard matter action. We still need to define the last bits of the theory, which are given by
\begin{eqnarray}
S_{\rm pre} &=& S_{\rm GR} +\frac{\Mpl^2}{2} \sum_{i=1}^4\int d^4x \mathcal{S}_i\,,\\
S_{\rm GR} &=& \frac{\Mpl^2}{2} \int d^4x N \sqrt\gamma [{}^{(3)}R+K^{ij}K_{ij}-K^2]\,,\\
\mathcal{S}_1 &=& -m^2 c_1 \sqrt{\tilde\gamma} (N+M\mathcal{K})\,,\\
\mathcal{S}_2 &=& -\frac12\,m^2 c_2 \sqrt{\tilde\gamma}(2N\mathcal{K}+M\mathcal{K}^2-M{\tilde\gamma}^{ij}\gamma_{ji})\,,\\
\mathcal{S}_3 &=& -m^2 c_3 \sqrt{\gamma} (M+N\mathfrak{K})\,,\\
\mathcal{S}_4 &=& -m^2 c_4 \sqrt{\gamma} N\,.
\end{eqnarray}

The theory has been defined on the unitary gauge (for the Stuckelberg fields), so we cannot make any further gauge choice. In what follows, we describe the background and perturbation analysis by using the following variables. For the scalar modes, the flat 3D metric is then written according to
\begin{equation}
    ds_3^2=\gamma_{ij} dx^i dx^j = [a(t)^2\,(1+2\zeta)\,\delta_{ij} + 2\partial_i\partial_jE ]\, dx^i dx^j\,,
\end{equation}
$a$ being the scale factor of the physical metric. Then we can define the lapse and shift vector as
\begin{eqnarray}
N&=&N(t)\,(1+\alpha)\,,\\
N_i&=&N(t)\,\partial_i\chi\,.
\end{eqnarray}
We also need to give
\begin{eqnarray}
\lambda &=& \lambda(t)+\delta\lambda\,,\\
\lambda^i &=& \frac{1}{a^2} \delta^{ij}\partial_j \delta\lambda_V\,.
\end{eqnarray}
Finally, we need to give all matter actions together with their own variables.

It is possible to show that on the background, $\lambda(t)=0$. In this case, the dynamics of MTMG imposes the following constraint
\begin{equation}
    \left( HX-H\,\frac{M}{N}+\frac{\dot{X}}{N} \right)  ({X}^{2}c_{{1}}+2\,Xc_{{2}}+c_{{3}} ) =0\,,\label{eq:branches}
\end{equation}
where we have defined the variable $X\equiv\tilde{a}/a$, being the ratio between the fiducial and the physical scale factors. The function $H=\dot{a}/(aN)$ is the Hubble parameter, and the coefficients $c_{1,2,3}$ are instead constant coefficients. On solving Eq.\ (\ref{eq:branches}), we can see that in this setup only two solutions are possible: 1) setting ${X}^{2}c_{{1}}+2\,Xc_{{2}}+c_{{3}} =0$, which defines the self-accelerating branch and leads to $X={\rm const.}$; 2) the normal branch for which $M=NX+\dot{X}/H$ where the time dependence of $X$ is still not imposed. Here onwards, we will only focus on the normal branch and, in particular, we fix the dynamics of $X$ so that $X=X_0={\rm const.}$

As a consequence of this choice, the background equations of MTMG read as follows:
\begin{eqnarray}
3\Mpl^2 H^2 &=& \sum_I\rho_I + \rho_{\rm MTMG}\,,\\
2\Mpl^2\,\frac{\dot{H}}{N} &=& -\sum_I(\rho_I+P_I)\,,\\
\frac{\dot{\rho}_I}{N}&=&-3\,H\,(\rho_I+P_I)\,,
\end{eqnarray}
where $\rho_{\rm MTMG} \equiv \frac {{m}^{2}\Mpl^2}{2} ( c_1 X_0^3+3c_2 X_0^2+3
c_3 X_0+c_4)$ is a constant. Here the sum over the index $I$ is over all the standard matter components. In this case, we can see that the background equations of motion exactly reduce to the ones of $\Lambda$CDM (endowed with an effective cosmological constant, i.e., the model is self-accelerating).

Let us now describe the dynamics of the perturbation equations of motion. We will follow here a procedure similar to the one introduced in \cite{Aoki:2020oqc} and \cite{Pookkillath:2019nkn,DeFelice:2020cpt}. We will have the perturbation equations of motion coming from the gravity sector (the ones for $\alpha$, $\chi$, $\zeta$, $E$, $\delta\lambda$ and $\delta\lambda_V$) together with the ones coming from each matter component. We then find the equations of motion for each of the fields we have. In the following, we will (and can) set $N(t)=a(t)$ as to have dynamics in terms of the conformal time. Although we cannot fix any gauge for the perturbations, we can still perform field redefinitions. In particular we will introduce the following ones:
\begin{eqnarray}
\alpha&=&\psi -\frac{\dot\chi}a +\frac1a\,\frac{d}{dt}\!\left[a\,\frac{d}{dt}\!\left( \frac{E}{a^2} \right)\right],\\
\zeta &=& -\phi -H\,\chi +aH \frac{d}{dt}\!\left( \frac{E}{a^2} \right).
\end{eqnarray}
These definitions correspond to the usual gauge-invariant definitions for the Bardeen potentials $\phi$ and $\psi$. Then we will use the Newtonian gauge invariant fields, $\phi$ and $\psi$, as to rewrite the modified Einstein equations. For each matter component, labeled with $I$, we also introduce gauge invariant combinations $\delta_I$'s and $\theta_I$'s (in place of the density contrast $\delta\rho_I/\rho_I$ and the scalar speed $u_{Ii}=\partial_i v_I$, respectively) as follows:
\begin{eqnarray}
\frac{\delta\rho_I}{\rho_I}&=&\delta_I -\frac{\dot{\rho}_I}{a\rho_I}\,\chi
+ \frac{\dot{\rho}_I}{\rho_I}\,\frac{d}{dt}\!\left( \frac{E}{a^2} \right),\\
v_I &=& -\frac{a}{k^2}\,\theta_I + \chi -a \,\frac{d}{dt}\!\left( \frac{E}{a^2} \right).
\end{eqnarray}
In terms of these variables the equations of motion for each matter component, for the lowest multipoles, read as follows:
\begin{eqnarray}
\dot{\delta}_I&=& 3a(w_I-c_{sI}^2) H \delta_I
- (1+w_I)\,(\theta_I - 3 \dot\phi)\,,\label{eq:dot_d_gen}\\
\dot{\theta}_I &=& aH (3c_{sI}^2 -1)\,\theta_I+{k}^{2}\psi
+{\frac {c_{sI}^2 k^2}{1+w_I}}\,\delta_I-{k}^{2}\sigma_I\,,\label{eq:dot_th_gen}
\end{eqnarray}
where $w_I\equiv P_I/\rho_I$, and $c_{sI}^2=\left(\frac{\partial p_I}{\partial\rho_I}\right)_s$ is the speed of propagation for each matter species (i.e., $\dot{p}_I/\dot{\rho}_I$, which vanishes for dust and equals 1/3 for photons). This result merely shows that the dynamical equations of motion for the matter components exactly coincide with the same ones in $\Lambda$CDM. This is not a surprise as modifications of gravity only enter in the gravity sector.

The equations of motion for $\delta\lambda_V$ impose
\begin{equation}
\chi=-\frac{\phi}{H}+\frac{\dot{E}}{a}-2HE\,.
\end{equation}
We can solve the equations of motion for $\chi$ in terms of $\delta\lambda$ (and other fields), and the equations of motion for $\alpha$ in terms of $E$. We can consider a linear combination of the equations of motion for $\zeta$ and $E$ in order to find a solution for $\delta\lambda_V$.

At this point, the equation of motion for $\delta\lambda$ gives
\begin{eqnarray}
\mathcal{E}_0 &\equiv& \left[ {\frac { ( Y\theta-2 ) {k}^{2}}{{a}^{2}}}+{\frac {9
\,\theta\,Y\Gamma}{2}} \right] \phi-3\sum_I\varrho_I\delta_I\nonumber\\
&+&{\frac {9aH ( Y\theta-2 )}{2{k}^{2}}}\,\sum_I\Gamma_I \theta_I=0\,,\label{eq:eq_E0}
\end{eqnarray}
where we have introduced $\rho_I=3\Mpl^2\varrho_I$, $P_I=3\Mpl^2 p_I$, $\Gamma_I\equiv \varrho_I+p_I$, $\Gamma=\sum_I\Gamma_I$, $Y=H_0^2/H^2$ and $\theta\equiv \frac {m^2X_0}{2H_0^2} ( c_1X_0^2+2c_2X_0+c_3 )$, which represents the only extra parameter which enters into the equations of motion in addition to the background effective cosmological constant. In the $\Lambda$CDM limit, whenever $|\theta Y|\ll1$, we recover the GR standard evolution.

The equation of motion for $E$, the last one available in the gravity sector, gives
\begin{eqnarray}
\mathcal{E}_1 &\equiv& \dot\phi+{\frac {3\,Y\theta\,a\Gamma\,\phi}{2\,H \left( Y\theta-2
 \right) }}+aH\psi+{\frac {3\,{a}^{2}}{{k}^{2} ( Y\theta-2 ) }}\sum_I\Gamma_I\theta_I\nonumber\\
 &=&0\label{eq:dot_phi_eq}\,.
\end{eqnarray}
On taking the time-derivative of $\mathcal{E}_0$ and removing $\dot\phi$ from $\mathcal{E}_1$, we find the following derived equation of motion:
\begin{eqnarray}
\mathcal{E}_2&\equiv&\psi+{\frac {9\,{a}^{2}}{2\,{k}^{2}}}\sum_I\Gamma_I\sigma_I-{\frac {9{a}^{2}\,\theta\,Y}{
{k}^{2} \left( 2\,Y\theta-4 \right) }}\sum_Ic_{sI}^2\varrho_I\delta_I\nonumber\\
&-& \left[ 
1+{\frac {3\theta\,Y\Gamma}{ \left( 2\,Y\theta-4 \right) {H}^{2}}}\right.\nonumber\\
&+&\left.
{\frac {27Y\theta a^2}{2\,{k}^{
2} ( 2-Y\theta)} \left( \sum_J c_{sJ}^2\Gamma_J -{\frac {{\Gamma}^{2
}}{2{H}^{2}}} \right) } \right] \phi\nonumber\\
&-&{\frac {27\,\theta\,Y\Gamma\,{a}^{3}}{4\,{k}^{4} ( Y\theta-2 ) H}}\sum_I\Gamma_I\theta_I=0\,,\label{eq:psi_eq}
\end{eqnarray}
where we have also used the matter equations of motion in order to replace $\dot\delta_I$, etc. In the limit $|\theta Y|\to0$, the above equations reduce to the standard GR form. We will use, as in GR, both $\mathcal{E}_1$ and $\mathcal{E}_2$ as the dynamical equations in the gravity sector. The fact that we do not have any additional dynamical equation of motion is a consequence of the fact that MTMG does not add any new dynamical degree of freedom in the theory.

Although the previous equations, never appeared before in the literature, completely define the behavior of gravity in MTMG, at any time and scale, we finish the description of the theory by considering the behavior of the theory under the influence of a single dust fluid at late times (i.e., neglecting radiation), in order to understand the effect of MTMG for the effective gravitational constant. Since this last step has already been performed in the literature (see e.g. \cite{Felice_2016,De_Felice_2017,Bolis:2018vzs}), this last calculation can then be considered to be a check of the calculations presented so far in this paper. Then on considering a dust fluid (mimicking CDM or late-time baryons), Eqs.\ (\ref{eq:dot_d_gen}) and~(\ref{eq:dot_th_gen}) reduce to
\begin{eqnarray}
\mathcal{E}_D &\equiv& {\dot\delta}_c+\theta_c-3\dot\phi=0\,,\label{eq:dot_d_eq}\\
\mathcal{E}_V &\equiv& {\dot\theta}_c+aH\theta_c-k^2\psi=0\,.\label{eq:dot_th_eq}
\end{eqnarray}
We can then solve Eq.\ (\ref{eq:psi_eq}) for $\psi$ in terms of $\phi$ and $\theta_c$ (dust having no shear, and vanishing $c_s^2$). On replacing, by using Eq.\ (\ref{eq:dot_phi_eq}), $\dot\phi$ into Eq.\ (\ref{eq:dot_d_eq}), we can then solve $\mathcal{E}_D=0$ for $\theta_c$ in terms of $\phi$ and ${\dot\delta}_c$. We can now replace, $\theta_c$, $\psi$ and $\dot\phi$ (coming from ${\dot\theta}_c$) inside Eq.\ (\ref{eq:dot_th_eq}), and solve $\mathcal{E}_V=0$ for $\phi$ in terms of ${\dot\delta}_c$  and ${\ddot\delta}_c$. Finally, on replacing $\phi$ and $\theta_c$ in Eq.\ (\ref{eq:eq_E0}), we find a closed second order differential equation for $\delta_c$. On evaluating this last equation for subhorizon scales, that is for $k/(aH)\gg 1$, we find
\begin{equation}
{\ddot\delta}_c + a H {\dot\delta}_c -\frac32\, \frac{G_{\rm eff}}{G_N} \,\varrho_c\, a^2 \,\delta_c = 0\,,
\end{equation}
where $\Mpl^2=(8\pi G_N)^{-1}$, $\Omega_c=\varrho_c(t)/H^2$, and
\begin{equation}
    \frac{G_{\rm eff}}{G_N}=\frac{2}{2-\theta Y}-\frac{3\theta Y \Omega_c}{(2-\theta Y)^2}\,,
\end{equation}
matching the result found, e.g.\ in \cite{De_Felice_2017}. The same dynamical equation holds also for the late-time baryonic fluid. Since we are going to use bounds on ISW-galaxy cross correlations, it is worthy to find the expression for the ISW-field, defined as $\psi_{\rm ISW}\equiv\phi+\psi$, in the subhorizon scales approximation. We find that
\begin{eqnarray}
\psi_{\rm ISW}&=&-\frac{3 H_{0}^{2} \Omega_{m0} }{k^{2}}\,\frac{\Sigma\,\delta_{m}}{a}\,,\\
\Sigma &\equiv&\frac{8-\left(4+3 \Omega_{m} \right)\theta Y}{2 \left(2-\theta  Y \right)^{2}} ,
\end{eqnarray}
where here $\Omega_m=\Omega_m(t)=\rho_{m}/(3\Mpl^2H^2)$, and $\Omega_{m0}=\Omega_m(z=0)$ (in the next sections, unless specified otherwise, $\Omega_m$ will be used to rewrite $\Omega_{m0}$). This result for $\psi_{\rm ISW}$ agrees with the one found in \cite{Bolis:2018vzs}.

Let us finally discuss the equations of motion for the gravitational waves propagating on this flat FLRW background (the vector modes do not show any deviation from GR). In fact, the contribution of MTMG to the tensor modes consists, by construction (see e.g.\ \cite{DeFelice_2015,Felice_2016}), of introducing a non-zero squared mass $\mu^2=H_0^2\theta$ to them. The equations of motion for the two tensor modes polarizations ($h_f$ labeled with $f\in\{{+},{\times}\}$) become
\begin{equation}
    {\ddot h}_f = -2\,\frac{\dot a}{a}\,{\dot h}_f - (k^2+\theta\,H_0^2\,a^2)\,h_f\,,
\end{equation}
where in the right hand side a source term can further be added.
\\

\begin{figure}
\begin{center}
\includegraphics[width=3.5in]{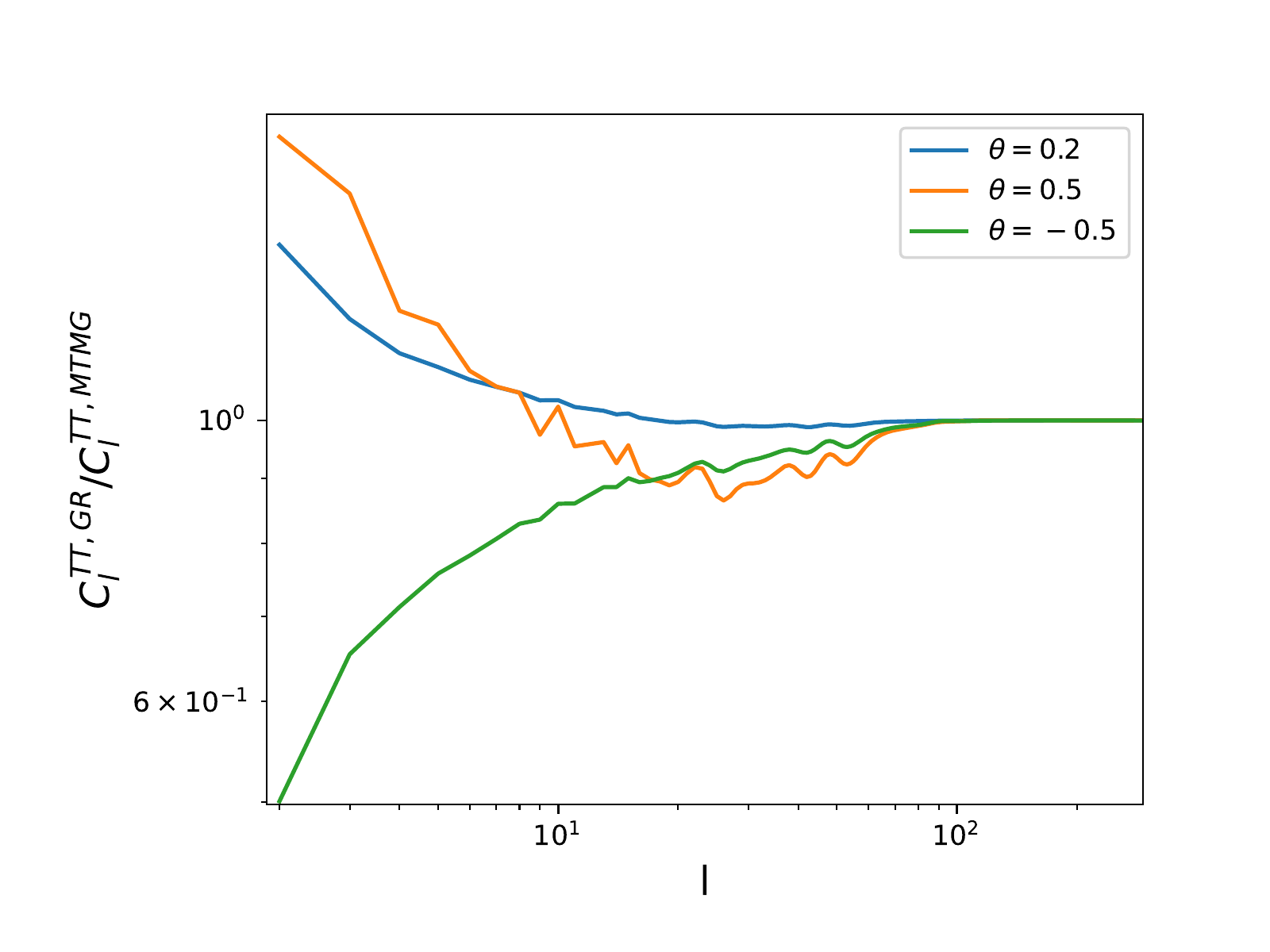}  
\caption{Deviations of the CMB TT power spectrum for $\Lambda$CDM from MTMG for some values of $\theta$.}
\label{Cl_TT_Pk_MTMG}
\end{center}
\end{figure}

\begin{figure*}
\begin{center}
\includegraphics[width=3.5in]{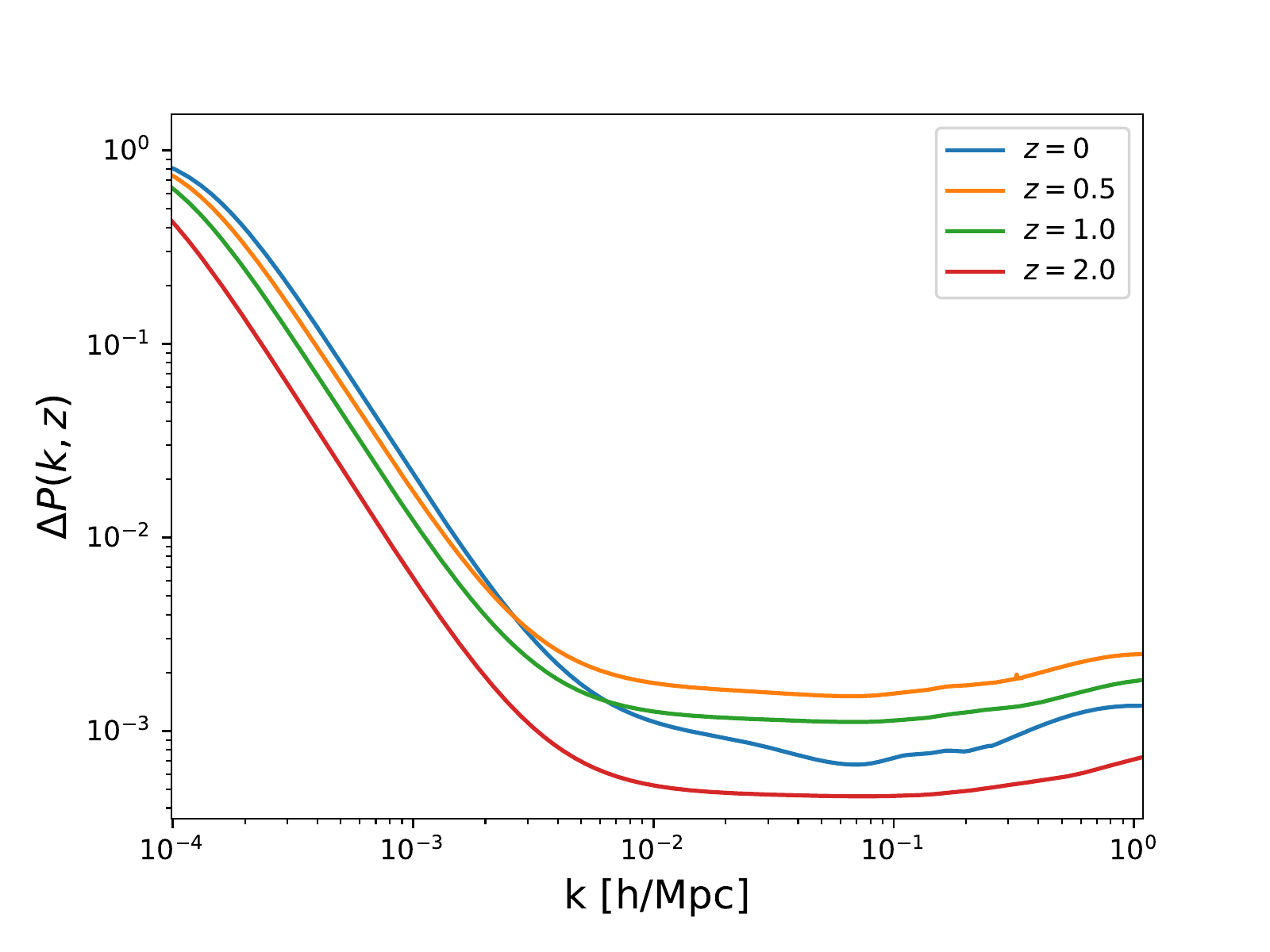}  
\includegraphics[width=3.5in]{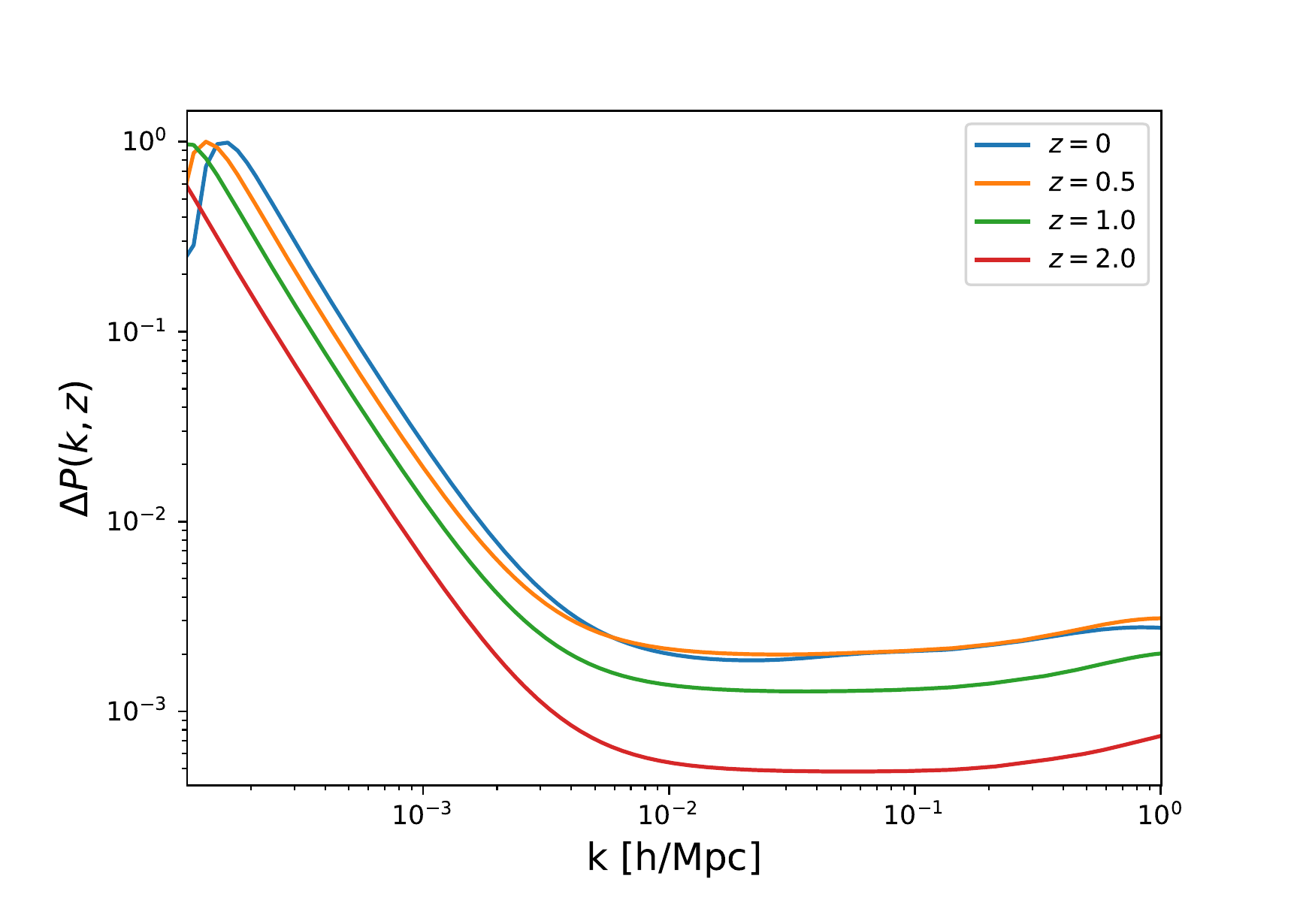}  
\caption{Left panel: Deviations of the matter power spectrum, $\Delta P(k,z) = |P(k,z)^{MTMG} - P(k,z)^{GR}|/(P(k,z)^{MTMG} + P(k,z)^{GR})$, for $\Lambda$CDM from MTMG with $\theta = -0.2$ for some values of $z$. Right panel: Same as in left panel, but assuming $\theta = 0.2$.}
\label{Pk_MTMG}
\end{center}
\end{figure*}

We implemented the model described above in \texttt{CLASS} code \cite{Blas:2011rf}. In Figure \ref{Cl_TT_Pk_MTMG}, we quantify the MTMG model affects on the CMB temperature power spectrum, i.e., $C_l^{TT}$, with respect to GR-$\Lambda$CDM prediction. We can see that increasing $\theta$ enhances the CMB temperature power spectrum at low $l$. This effect is well known for dark energy and MG models, and it is due to the integrated Sachs-Wolfe effect, which affects the CMB spectrum at low $l$, but has no significant effect at large scales. For scales larger than $l > 100$, we note $C_l^{GR, TT} \approx C_l^{MTMG, TT}$. In Figure \ref{Pk_MTMG}, we show the difference on the $P(k, z)$ at four different redshift values on the scales $k \in [10^{-4}, 1]$ h/Mpc. These $z$ values are chosen because our RSD sample  covers the range $z \in [0.02, 1.94]$. For a quantitative example, taking $\theta = -0.2$, we find that the difference with respect to $\Lambda$CDM is $\sim 25$\% at $k = 10^{-3}$ h/Mpc, while for $k > 10^{-2}$ h/Mpc the difference is $< 0.2$\%, for all $z$ values. For the case with $\theta = 0.2$, we note that for $k > 10^{-3}$ h/Mpc, we have a similar difference as in the previous case, but with different $P(k, z)$ behavior at different $z$ values, as expected, since the dynamics  depends on the signal in $\theta$. For scales $k < 10^{-3}$, significant deviations are noted. It is important to mention that on very large scale, the cosmic variance effects can be significant. In these simulations, the predictions of the non-linear effects are performed using the HMcode code \cite{Mead_2015}. Non-linear dynamics of the MTMG framework using N-body simulations were investigated in detail in \cite{Hagala_2021}.


\begin{table*}
    \centering
    \caption{Constraints at 95\% CL on $\Omega_m$, $\sigma_8$, $S_8$ and $\theta=\mu^2/H_0^2$, inferred from different data set combinations, in the MTMG model.}
    \label{tab:main_results_RSD}
    \begin{tabular}{ccccc}
        \hline
        \hline
		data set & $\theta$ & $\Omega_m$ & $\sigma_8$ & $S_8$  \\ 
		\hline
		RSD & $-5.2^{+7.1}_{-4.8}$ & $0.37^{+0.17}_{-0.16}$ &  $0.808^{+0.086}_{-0.084}$ &  $0.89^{+0.19}_{-0.20}$ \\ 
		RSD + BAO + Pantheon & $-1.8^{+3.3}_{-4.4}$ & $0.305^{+0.041}_{-0.039}$ & $0.796^{+0.091}_{-0.073}$ & $0.802^{+0.10}_{-0.096}$ \\ 
		RSD + BAO + Pantheon + ISW& $-0.12^{+0.28}_{-0.26}$ & $0.293^{+0.018}_{-0.018}$ & $0.775^{+0.055}_{-0.055}$ & $0.766^{+0.057}_{-0.055}$ \\
		\hline
    \end{tabular}
\end{table*}

\begin{figure*}
\begin{center}
\includegraphics[width=3in]{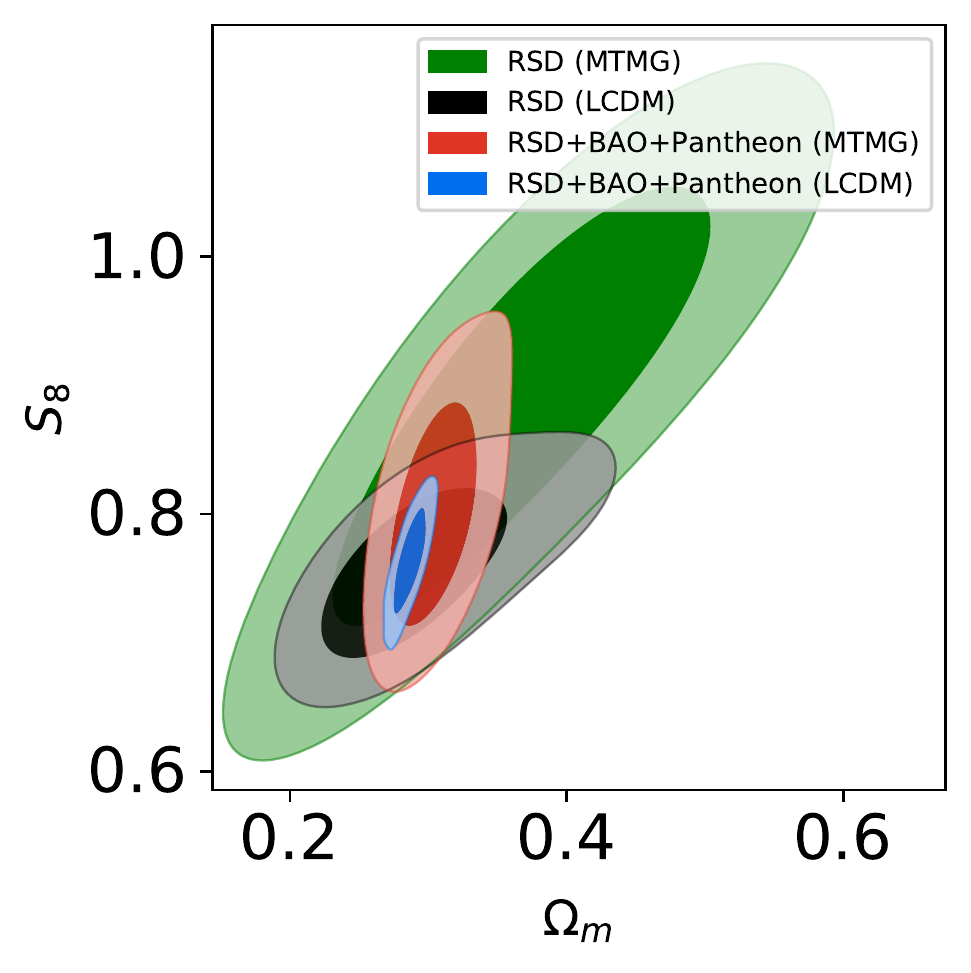}
\includegraphics[width=3in]{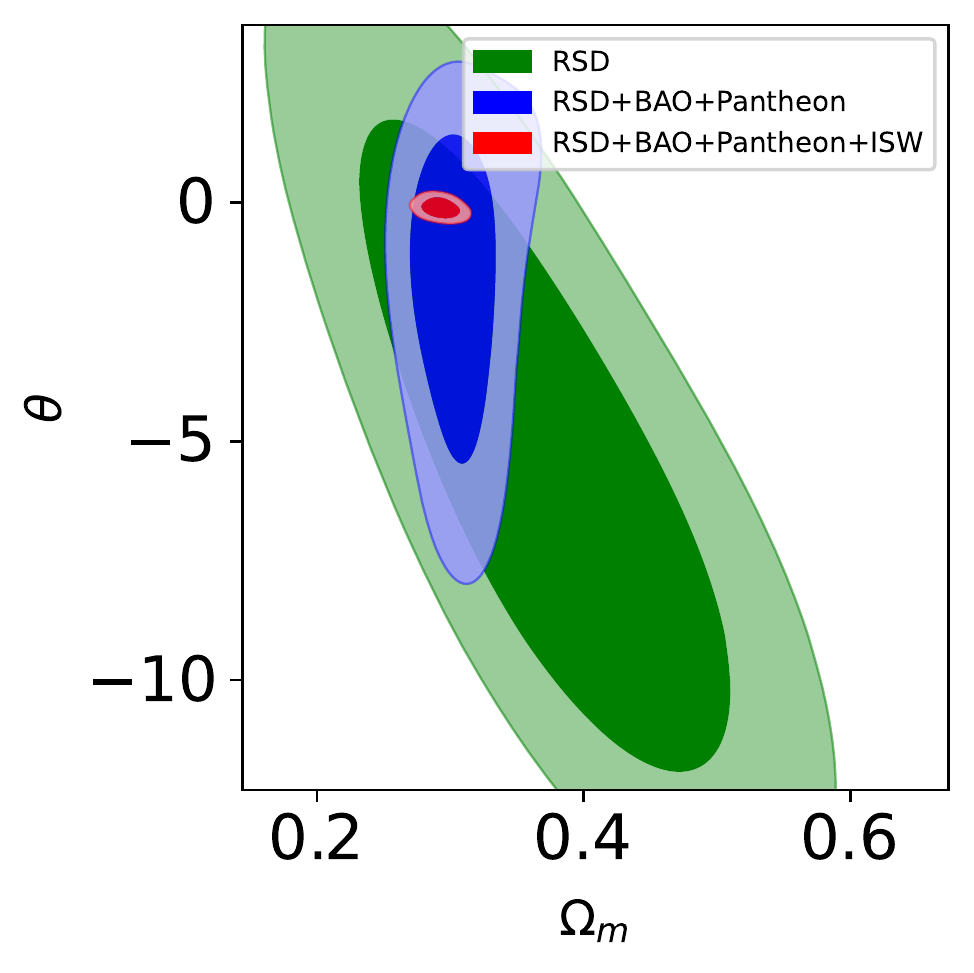} 
\caption{Left panel: 68\% and 95\% confidence levels on the $S_8-\Omega_m$ plane from various data sets in $\Lambda$CDM and MTMG models. Right panel: 68\% and 95\% confidence levels on the $\theta-\Omega_m$ plane from various data sets in MTMG model. GR is recovered for $\theta = 0$.}
\label{S8}
\end{center}
\end{figure*}

\section{Data and Methodology}
\label{data}

In this section, we present the data sets and methodology used to obtain the observational constraints on the model parameters by performing Bayesian MCMC analysis. In order to constrain the parameters, we use the following data sets. \\

\noindent\textbf{CMB:} We use the full Planck-2018 \cite{2020} CMB temperature and polarization data which comprise of the low-l temperature and polarization likelihoods at $l \leq 29$, temperature (TT) at $l \geq 30$, polarization (EE) power spectra, and cross correlation of temperature and polarization (TE) while also including the Planck-2018 CMB lensing power spectrum likelihood \cite{2020_lensing} in our analysis. 
\\

\noindent\textbf{RSD:} The growth rate data comprise of $f(z)\sigma_8(z)$ measurements from a variety of redshift space distortion surveys. The current measurements of $f\sigma_8(z)$ come from a plethora of different surveys with different assumptions and systematics, thus an approach to study the statistical properties and robustness of the data is imperative. In~\cite{Sagredo:2018ahx}, an internal robustness analysis was used to validate a subset of $f\sigma_8(z)$ measurements. In our analysis, we use the compilation of the $f\sigma_8(z)$ measurements presented in Table I in \cite{Sagredo:2018ahx}, which are minimally affected by systematic-contaminated data-points.  We refer this data set to as the data set of RSD measurements. To build the likelihood function, we follow the same methodology as presented in \cite{Arjona_2020}. It is important to highlight that the RSD sample is still fiducial to some cosmology, for example fixed on a $\Lambda$CDM baseline, different assumptions on the reference value of $\Omega_m$, and non-linearities modeling of which starts to play an important role on smaller scales and at later epochs. These  points should be taken into account to accurately estimate the cosmological parameters. Our sample is minimally affected by systematic-contaminated $f\sigma 8$ data points, checked through Bayesian model comparison framework described in \cite{Sagredo:2018ahx},  identifying potential outliers as well as subsets of data affected by systematics or new physics. \\

\noindent\textbf{BAO:} The Baryon Acoustic Oscillations (BAO) provide an important cosmological probe, which can trace expanding spherical wave of baryonic perturbations from acoustic oscillations at recombination time through the
large-scale structure correlation function. In this work, we consider the most recent BAO data compilation comprised of the $D_V(z)/r_d$, $D_M(z)/r_d$, and $D_H(z)/r_d$ measurements compiled in Table~3 in ~\cite{Alam_2021}, regarding BAO-only data.\\ 

\noindent\textbf{Pantheon:} The Supernovae Type Ia  have traditionally been one of the most important astrophysical tools in establishing the so-called standard cosmological model. For the present analysis, we use the Pantheon compilation, which consists of 1048 SN Ia distributed in the redshift range $z \in [0.01, 2.3]$~\cite{Scolnic:2017caz}. \\



\noindent\textbf{ISW:} The late-time Integrated Sachs-Wolfe effect (ISW) on the CMB is an effect imprinted in the angular pattern of the CMB in the presence of a time varying cosmological gravitational potential, which can be due to a non-flat universe \cite{Kinkhabwala_1999}, as well as for a flat one in the presence of dark energy or modified gravity theories \cite{Giacomello_2019,Bolis_2018,Song_2007}. A non-zero ISW necessarily implies the presence of a generating physical source for the accelerated expansion of the universe at late times. For the present analysis, we use the cross correlation of the CMB with galaxy surveys that derive constraints on the ISW as obtained in \cite{St_lzner_2018}. We use the five catalogs of extragalactic sources as presented in \cite{St_lzner_2018}, namely, the 2MASS Photometric Redshift catalog, the WISE × SuperCOSMOS photo-z catalog, the NRAO VLA Sky Survey radio sources catalog, the SDSS DR12 and SDSS DR6 QSO photometric catalogs.
\\

We use the Metropolis-Hastings mode in \texttt{CLASS}+\texttt{MontePython} code \cite{Blas:2011rf,Audren:2012wb,Brinckmann:2018cvx}  to derive the constraints on cosmological parameters using various data combinations from the data sets described above, ensuring a Gelman-Rubin convergence criterion of $R - 1 < 10^{-3}$. In what follows, we describe our main results.

\section{Results and discussions}
\label{results}

We divide the analysis into two parts. First, we consider RSD and RSD + BAO + Pantheon data, that is, the growth data and its combination with the geometric data. Thus, we can quantify the dynamics of the model and the constraints on $\theta$ without CMB data influence. It is important to check  the prediction of the model about $\sigma_8$ or $S_8$ in the absence of the CMB data. Then, we add ISW information to these data. In the second round of analysis, we analyze the model with CMB data, and discuss the potential of the model to solving and/or alleviating the $H_0$ and $S_8$ tensions.

\begin{table*}\caption{Summary of the 1$\sigma$ constraints on  the baseline parameters of the MTMG scenario from CMB and its combination with several other data sets. In the last column, the Full Joint Analysis means CMB + BAO + Pantheon + RSD + ISW. The parameter $H_{\rm 0}$ is measured in the units of km s${}^{-1}$ Mpc${}^{-1}$.}
\begin{tabular} {|l|l|l|l|l|}\hline
\centering
 Parameter &  CMB & CMB + BAO + Pantheon &  CMB + BAO + Pantheon + RSD & Full Joint Analysis \\
\hline
$10^{2}\omega_{b }$ &  $2.242\pm 0.015  $ & $2.245\pm 0.013$ & $2.252\pm 0.013 $ & $2.247\pm 0.013$ \\

$\omega_{\rm cdm }  $ &  $0.1195\pm 0.0012$ & $0.11903\pm 0.00090  $ &  $0.11819\pm 0.00090 $ & $0.11886\pm 0.00095$ \\

$100\theta_{\rm s}  $   & $1.04195\pm 0.00029 $ &  $1.04197\pm 0.00028  $ & $1.04202\pm 0.00028  $ & $1.04197\pm 0.00028$ \\

$\ln10^{10}A_{\rm s }$ &  $3.042\pm 0.014   $ &  $3.042\pm 0.014  $ &  $3.035\pm 0.014  $ & $3.036\pm 0.015$ \\

$n_{\rm s } $ &  $0.9663\pm 0.0044 $ &  $0.9672\pm 0.0036 $ &  $0.9692\pm 0.0039$ & $0.9681\pm 0.0037$ \\

$\tau_{\rm reio} $ &  $0.0538\pm 0.0071$ &  $0.0542\pm 0.0071$ &  $0.0520\pm 0.0070$ & $0.0517\pm 0.0075$ \\

$\theta$ &  $0.19^{+0.14}_{-0.10} $ &   $0.21^{+0.13}_{-0.10} $ &  $0.26^{+0.14}_{-0.10}$ & $0.129^{+0.13}_{-0.088}$ \\

$\Omega_{\rm m}$ &   $0.3115\pm 0.0077           $ &   $0.3089\pm 0.0054          $&   $0.3038\pm 0.0054  $ & $0.3018\pm 0.0056$ \\

$H_0$ &  $67.65\pm 0.57   $ &  $67.84\pm 0.40 $ &  $68.22\pm 0.42$ & $68.44\pm 0.43$ \\

$\sigma_8$ &  $0.8071\pm 0.0064 $ &  $0.8057\pm 0.0065  $ &  $0.8001\pm 0.0060 $ & $0.8169\pm 0.0068$ \\
$S_8$ &  $ 0.822\pm 0.014 $ &  $ 0.818\pm 0.011  $ &  $ 0.805\pm 0.010 $ & $ 0.819\pm0.011 $ \\
\hline
\end{tabular}\label{cmb_results}
\end{table*}

\begin{figure*}
\begin{center}
\includegraphics[width=3in]{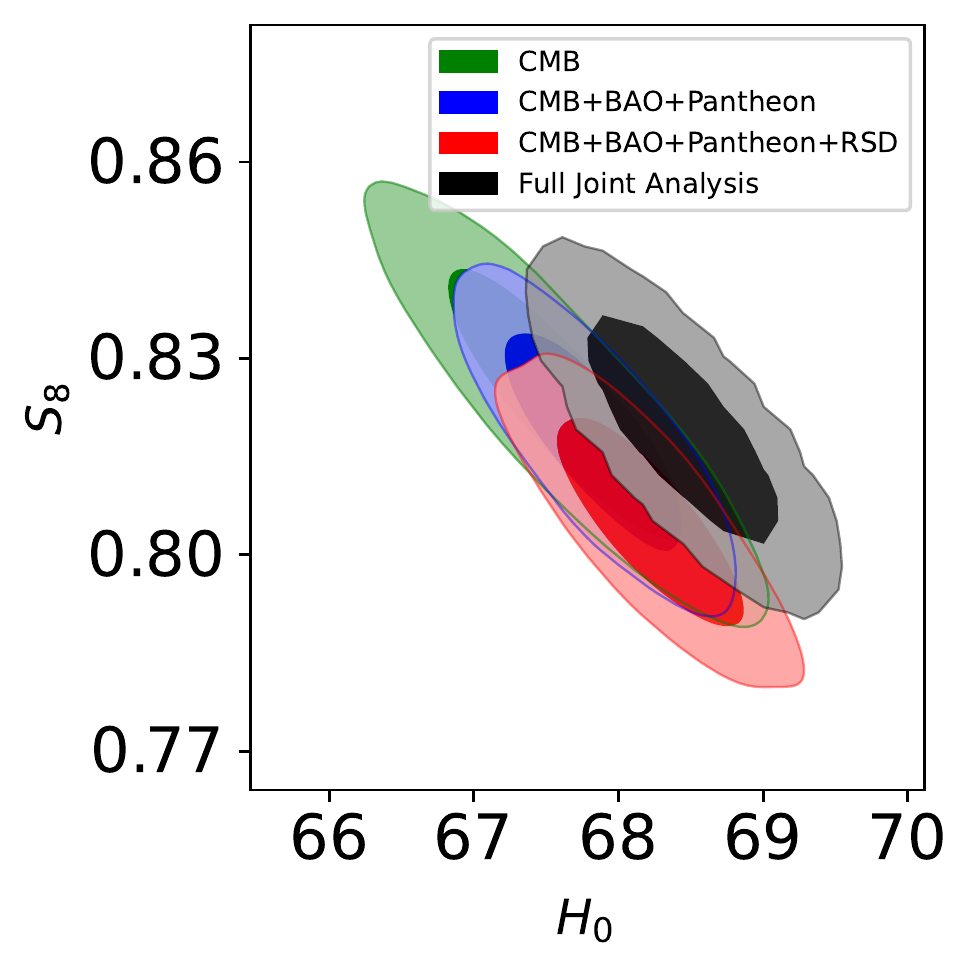}
\includegraphics[width=3in]{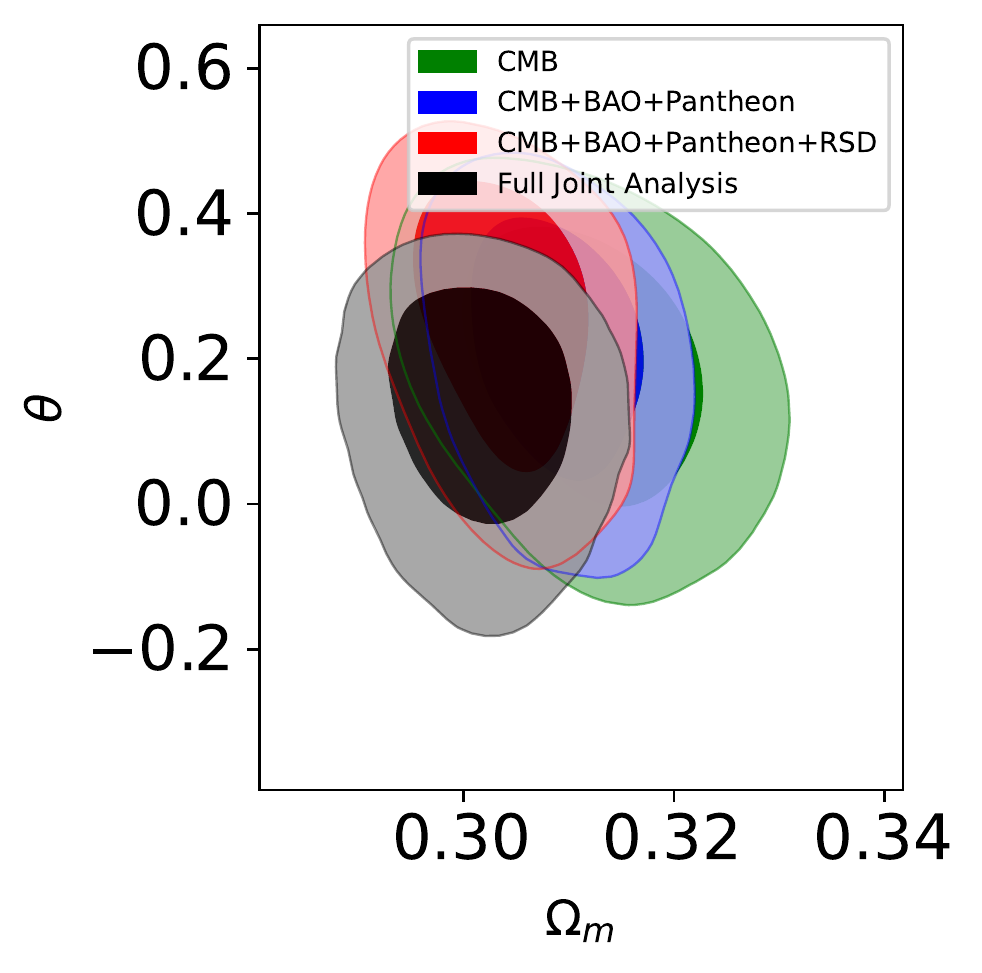} 
\caption{Left panel: 68\% and 95\% CLs in the $S_8-H_0$ plane from various data sets in the MTMG model. Right panel: Same as the left panel, but in the $\theta-\Omega_m$ plane. GR is recovered for $\theta = 0$.} 
\label{theta_full}
\end{center}
\end{figure*}

\subsection{Analysis with the growth, geometric and ISW data}

Model-wise, we consider $\Lambda$CDM+$\theta$ model baseline, spanned by the 5 parameters: the Hubble constant $H_0$ or equivalently the reduced Hubble constant $h \equiv H_0/(100\,{\rm km}\,{\rm s}^{-1}\,{\rm Mpc}^{-1})$, the physical baryon density $\omega_b \equiv \Omega_bh^2$, the physical cold dark matter density $\omega_c \equiv \Omega_ch^2$, $\sigma_8$ and $\theta$, quantifying deviations from GR. The matter density parameter today $\Omega_m$ is derived by $\Omega_m = \Omega_b+\Omega_c$. Another important derived parameter is $S_8 \equiv \sigma_8\sqrt{\Omega_m/0.3}$.  To constrain the physical baryon density, we adopt a Gaussian prior on $\omega_b$ from Big Bang Nucleosynthesis (BBN): $100\omega_b =  2.233 \pm 0.036$~\cite{Mossa:2020gjc}. In order that the tensor modes not to develop instability whose time scale is shorter than the age of the universe, we impose the prior $\theta \geqslant - 10$. In the limit $\theta = 0$, we recover the evolution equation for the perturbations as in GR. Note that the $\Lambda$CDM and MTGM models are indistinguishable at the background evolution. Thus, the data combination BAO + Pantheon is indistinguishable for both scenarios. We use BAO + Pantheon to break the statistical degeneracy on $\Omega_m$, that may be present when considering RSD data only. Since both $\Lambda$CDM and MTGM models can predict different constraints on $\Omega_m$ in light of only the RSD data, breaking this background degeneracy with BAO + Pantheon can significantly improve the results. Finally, we also consider ISW sample data. Table \ref{tab:main_results_RSD} summarizes the main results of our statistical analyses. 

Figure \ref{S8}, on the left panel, shows the parameter space in the $S_8 - \Omega_m$ plane at 68\% and 95\% CL with various data sets in $\Lambda$CDM and MTMG models. First, we note that, in direct comparison with the constraints provided with $\Lambda$CDM, the MTMG model can provide an enlargement in both $S_8$ and $\Omega_m$ estimations. The right panel shows the regions at 68\% and 95\% CL in the $\theta-\Omega_m$ plane with RSD and RSD + BAO + Pantheon data in the MTMG model. We find $\theta=-5.2^{+7.1}_{-4.8}$, $\theta=-1.8^{+3.3}_{-4.4}$ and $\theta=-0.12^{+0.28}_{-0.26}$ at 95\% CL from RSD, RSD + BAO + Pantheon and RSD + BAO + Pantheon + ISW data, respectively. The addition of the ISW data improves the constraints on $\theta$ because $\theta$ can significantly affect the late-time ISW effect \cite{Bolis:2018vzs}. So, ISW data improve $\theta$, and consequently by correlation, improvements on the other parameters in the baseline are also reached, which are already well constrained from other data-sets. The joint analysis with RSD + BAO + Pantheon + ISW data provides robust and accurate constraints on the theory using only late time probes. As we will show in the next section, the constraints are further improved with the addition of CMB data. All these constraints are consistent with the GR prediction, i.e., $\theta=0$ even at 68\% CL. For comparison, assuming $\Lambda$CDM cosmology, from RSD data, we find $\Omega_m=0.293^{+0.10}_{-0.083}$ and $S_8=0.758^{+0.082}_{-0.073}$. From the joint analysis with RSD + BAO + Pantheon data, we find $\Omega_m=0.286^{+0.015}_{-0.015}$ and $S_8=0.765^{+0.055}_{-0.054}$. We can clearly see that the addition of BAO + Pantheon improves the constraint by breaking down the degeneracy on $\Omega_m$, in both models, i.e., $\Lambda$CDM and MTMG. As we will see in the next section, in MTMG model, there is no tension in the $S_8 - \Omega_m$ plane, when analyzed from CMB and RSD data. Therefore, MTMG brings an agreement between these data.

\subsection{MTMG in the light of CMB data}

For the first time, here we find the constraints on the MTMG model from the full Planck-CMB data set alone and its combination with several other data. The baseline seven free parameters set of the MTMG model is given by:
\begin{equation*}
\label{baseline1}
\mathcal{P}= \Big\{ \omega_{\rm b}, \, \omega_{\rm cdm}, \, \theta_s, \,  \ln(10^{10} A_s), \, n_s, \, \tau_{\rm reio}, \,   \theta \Big\},
\end{equation*}
where the first six parameters are the baseline parameters of the standard $\Lambda$CDM model, namely: $\omega_{\rm b}$ and $\omega_{\rm cdm}$ are respectively the dimensionless densities of baryons and cold dark matter; $\theta_s$ is the ratio of the sound horizon to the angular diameter distance at decoupling; $A_s$ and $n_s$ are respectively the amplitude and spectral index of the primordial curvature perturbations, and $\tau_{\rm reio}$ is the optical depth to reionization. As commented before, the parameter $\theta$ quantifies deviations from GR induced by the MTMG framework.

Table \ref{cmb_results}  displays summary of the our statistical analyses using the full Planck-CMB data set and its combination with several other data sets. In the last column, the Full Joint Analysis stands for CMB + BAO + Pantheon + RSD + ISW. Figure \ref{theta_full}, on the left panel (right panel), shows the parameter space in the $S_8 - H_0$ ($\theta - \Omega_m$) plane, at 68\% and 95\% CL from Planck-CMB data set and its combination with other data.

\begin{figure*}
\begin{center}
\includegraphics[width=3in]{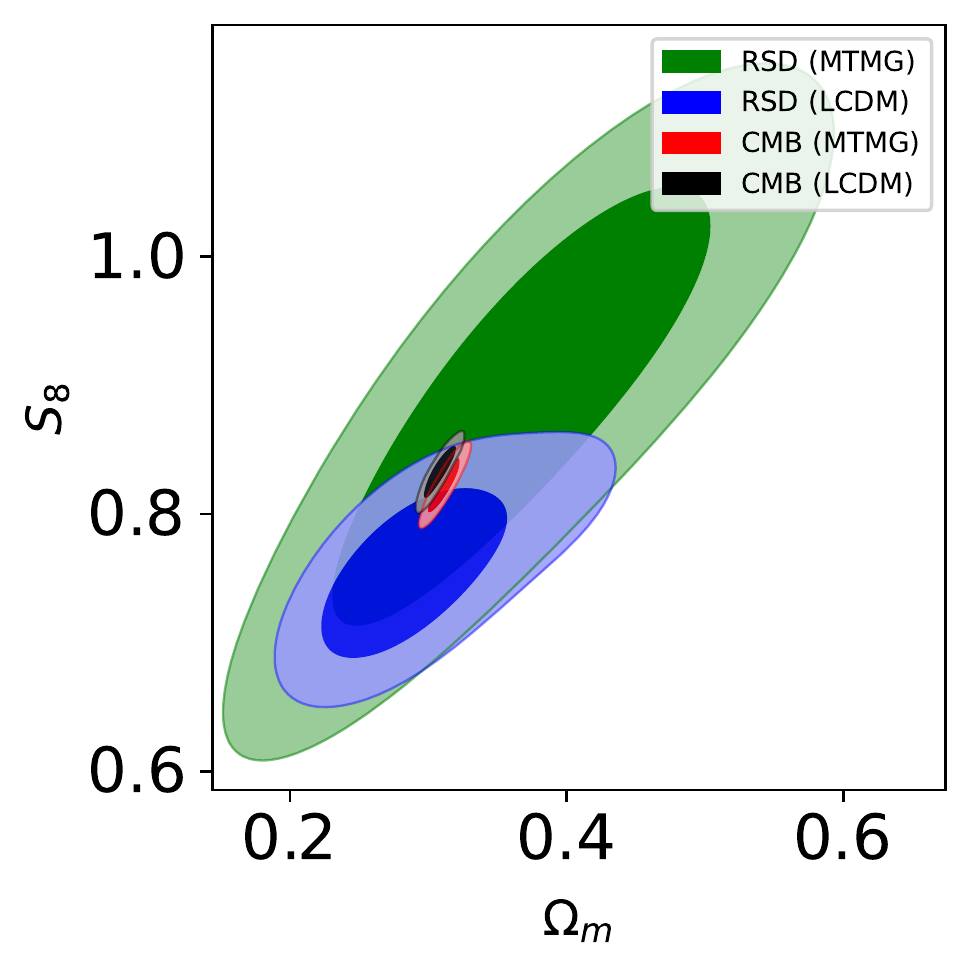}
\includegraphics[width=3in]{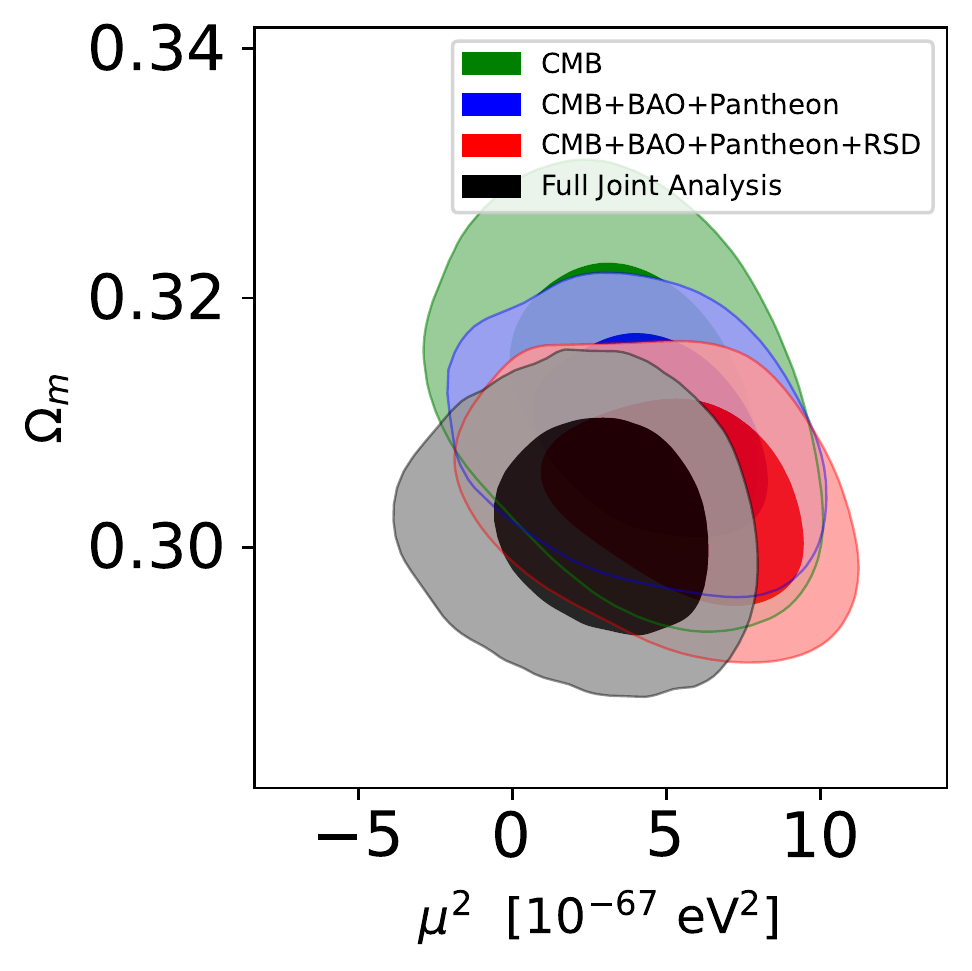}
\caption{Left panel: 68\% and 95\% CLs on the parametric space $S_8-\Omega_m$ from CMB and RSD data sets under the $\Lambda$CDM and MTMG model assumptions. Right panel: 68\% and 95\% CLs on the parametric space $\mu^2-\Omega_m$ from several different data sets.}
\label{S8_tension}
\end{center}
\end{figure*}

In all the analyses carried out here, we note that $\theta$ can be non-null at 68\% CL, but consistent with $\theta=0$ at 95\% CL. Thus, we do not find any significant evidence for deviations from GR. In the Full Joint Analysis, we obtain $\theta=0.25^{+0.16}_{-0.10}$ at 68\% CL. The constraints on MTMG using the CMB data alone are well consistent with the $\Lambda$CDM baseline. Despite $\theta$ being non-null at 68\% CL, all other parameters do not show significant deviation from $\Lambda$CDM baseline. When BAO + Pantheon and BAO + Pantheon + RSD are added, we notice a minor shift of $\Omega_m$ to low values and $H_0$ to high values. This behavior is clear in Fig. \ref{theta_full}. We find that RSD and CMB  data are not in tension with each other in MTMG, see Fig. \ref{S8_tension}. On the other hand, these data are known to be in tension ($\sim 3\sigma$) within $\Lambda$CDM \cite{nunes2021arbitrating}. Thus, we can combine CMB and RSD in MTMG scenario. In other words, there is no tension in the $S_8-\Omega_m$ plane for the MTMG model.

\begin{figure*}
\begin{center}
\includegraphics[width=3in]{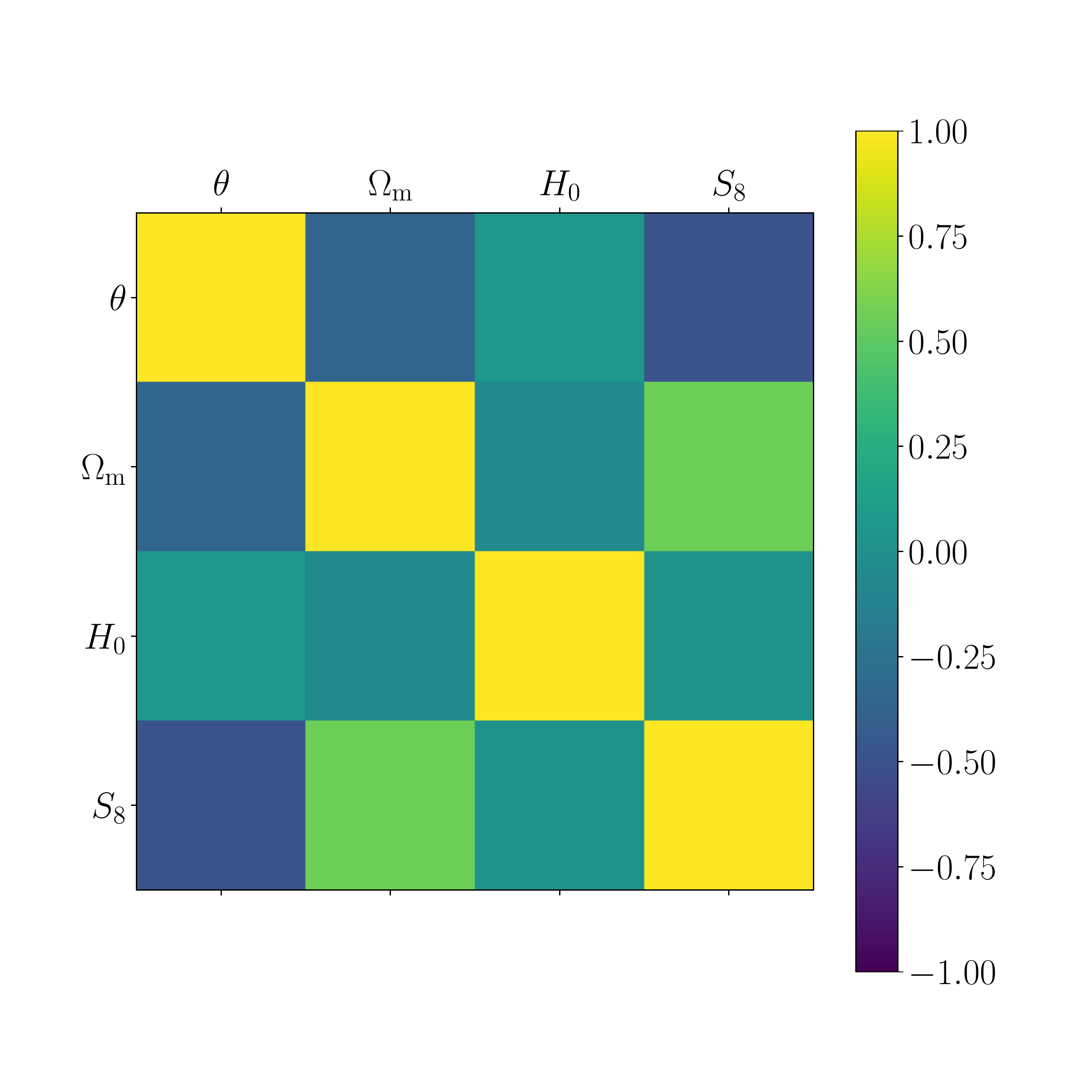}
\includegraphics[width=3in]{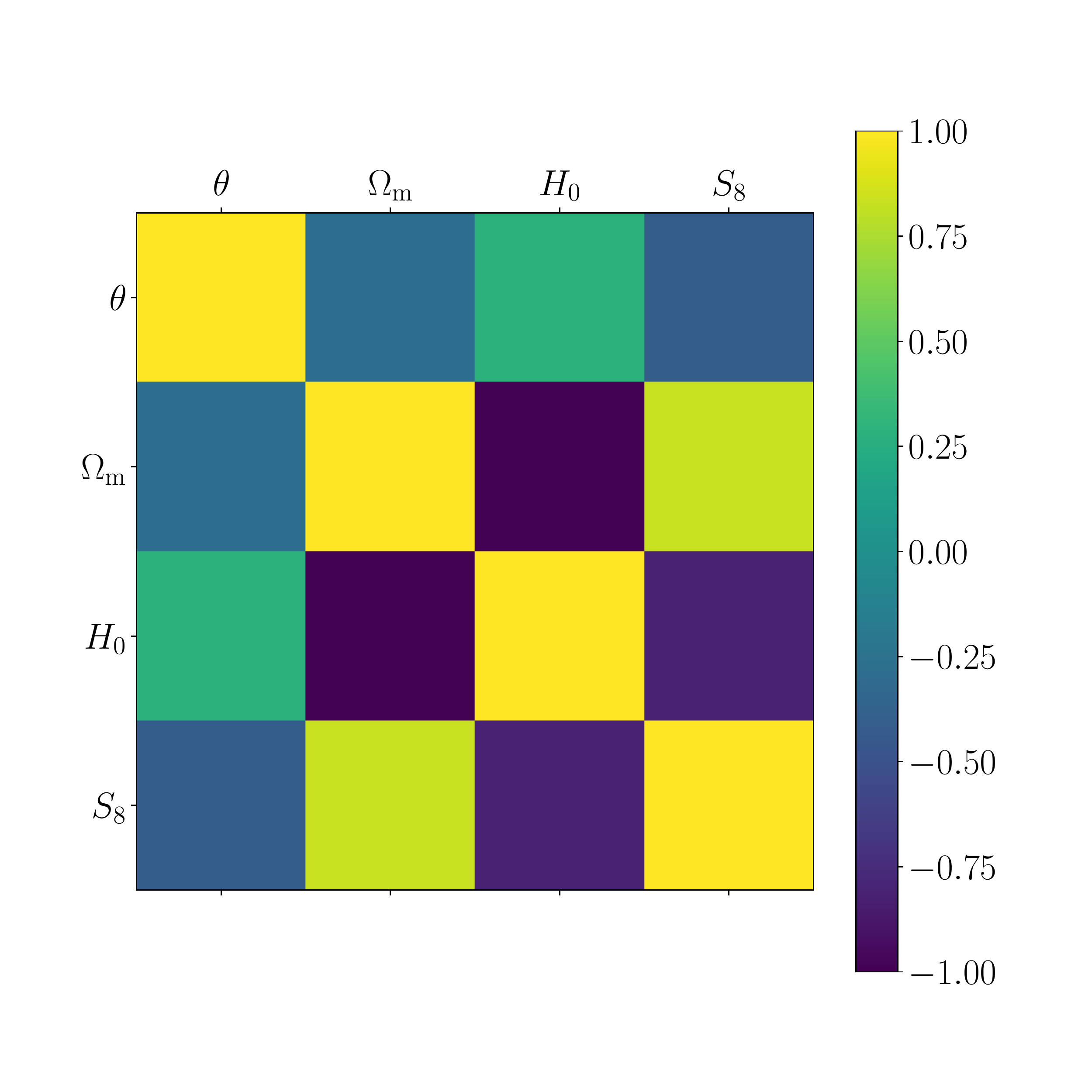} 
\caption{Left panel: Scatterplot quantifying the correlations for some parameters of interest in MTMG model from the  RSD + BAO + Pantheon + ISW data. Right panel: Same as the left panel, but for CMB + RSD + BAO + Pantheon + ISW data.}
\label{corr_plot}
\end{center}
\end{figure*}

We know that MTMG affects the CMB spectrum predominantly at low $l$, and practically no effects are observed at high $l$. Thus, we also consider ISW data in combination with the data-sets described above. We find $H_0=68.44 \pm 0.43$ km s$^{-1}$Mpc$^{-1}$ from CMB + BAO + Pantheon + RSD + ISW data. This constraint minimizes the $H_0$ tension up to 3.5$\sigma$ compared to cosmological model-independent local measurement of $H_0$ from the HST observations \cite{Riess_2019}. From our joint analysis, we have $\Omega_{\rm m}=0.301\pm 0.0056$ and $\sigma_8=0.8169\pm 0.0068$, which represents a deviation to low (high) values on $\Omega_{\rm m}$ ($\sigma_8$), respectively, compared to the CMB only constraints. 

Figure \ref{corr_plot} shows a scatterplot  for the correlations between the parameters of interest in our model from the RSD + BAO + Pantheon + ISW and CMB + RSD + BAO + Pantheon + ISW data. We note that in both cases, $\theta$ exhibits negative, positive and negative correlations with $\Omega_{m}$, $H_0$ and $S_8$, respectively. We found similar correlations in other analyses as well.  Despite all the constraints on $\theta$, from the different analyses performed here be statistically consistent with each other, it is interesting to note that $\theta$ shows a best fit preference to $\theta < 0$ from RSD + BAO + Pantheon + ISW, and $\theta > 0$ when we add the CMB data. We know that $\theta < 0$ causes a suppression on the amplitude of the matter perturbations. This suppression behavior is necessary to fit well, for instance, with the RSD data and its combination with BAO + Pantheon sample. On the other hand, in light of these data,  the parameter $\theta$ is degenerate with the other baseline parameters. Thus, it is expected that in the presence of RSD data with BAO + Pantheon + ISW data, preference exists for $\theta < 0$ values while not excluding the possibility of $\theta > 0$ values. Next, the addition of CMB data is expected to yield a tight constraint on $\theta$. We find in this case that the correlation of $\theta$ with $\Omega_{m}$, $H_0$ and $S_8$, increases. It is quantified in Fig. \ref{corr_plot} comparing both panels. That increase especially occurs on $\Omega_{m}$ and $H_0$, where the correlation with $H_0$ increases 400\%. Note that $H_0$ is strongly constrained using CMB data, in MTMG model, $H_0$ is constrained to 0.6\% accuracy. The global constraint on $\theta$ is also  very tight, and the best fit on $\theta$ changes the signal due to the correlation change in the analysis RSD + BAO + Pantheon + ISW to CMB + RSD + BAO + Pantheon + ISW. That is caused by the addition of the CMB data. We also note from the joint analysis CMB + RSD + BAO + Pantheon + ISW that $\theta>0$ at 68\% CL. Changes in the other baseline parameters are understandable looking in Figure \ref{corr_plot}.

It is interesting to note that CMB data can break the degeneracy present in RSD and BAO + Pantheon + RSD data, providing a tight constraint on $\theta$ when all these data are combined. It is clear from a comparison between the tables \ref{tab:main_results_RSD} and \ref{cmb_results}. Using the relation $\mu^2 = H_0^2 \theta$, and constraints on $H_0$, we can infer direct constraints on the graviton mass. Figure \ref{S8_tension} on the right panel shows the constraint on the graviton mass squared from CMB alone, CMB + BAO + Pantheon, CMB + BAO + Pantheon + RSD and CMB + BAO + Pantheon + RSD + ISW. If we take an inference with prior $\mu^2 \ge 0$, we find the bound: $\mu \lessapprox 8.8 \times 10^{-67}$ ${\rm eV}^2$, $\lessapprox 9.2 \times 10^{-67}$ ${\rm eV}^2$ eV, $\lessapprox 9.9 \times 10^{-67}$ ${\rm eV}^2$ eV, $\lessapprox 7.2 \times 10^{-67}$ ${\rm eV}^2$ eV at 95\% CL from CMB alone, CMB + BAO + Pantheon, CMB + BAO + Pantheon + RSD and CMB + BAO + Pantheon + RSD + ISW, respectively. Without loss of generality, this inference on $\mu$ can be performed for the other data sets. We chose to perform only using the most accurate measurements.

\begin{figure*}
\begin{center}
\includegraphics[width=3in]{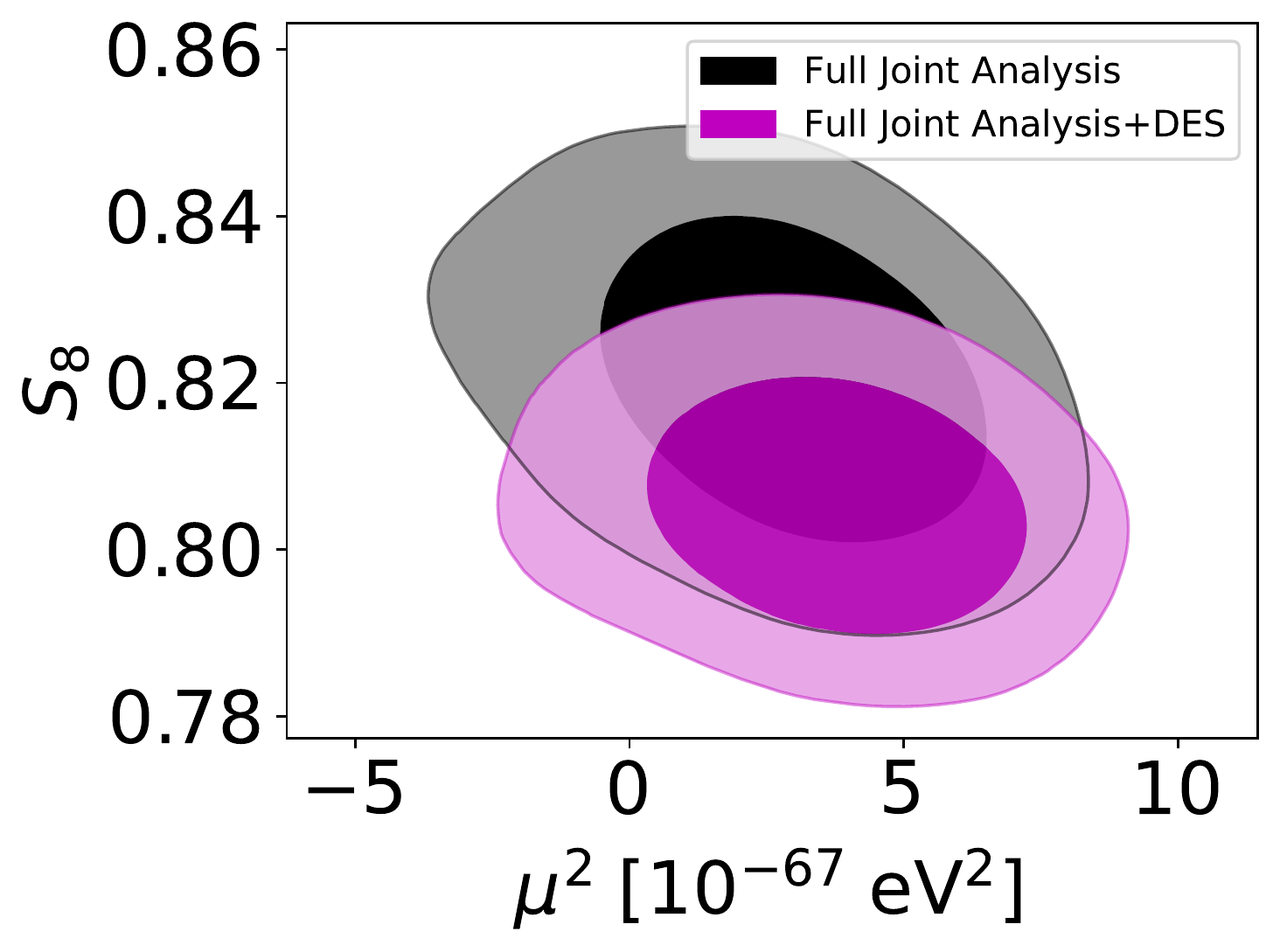}
\includegraphics[width=3in]{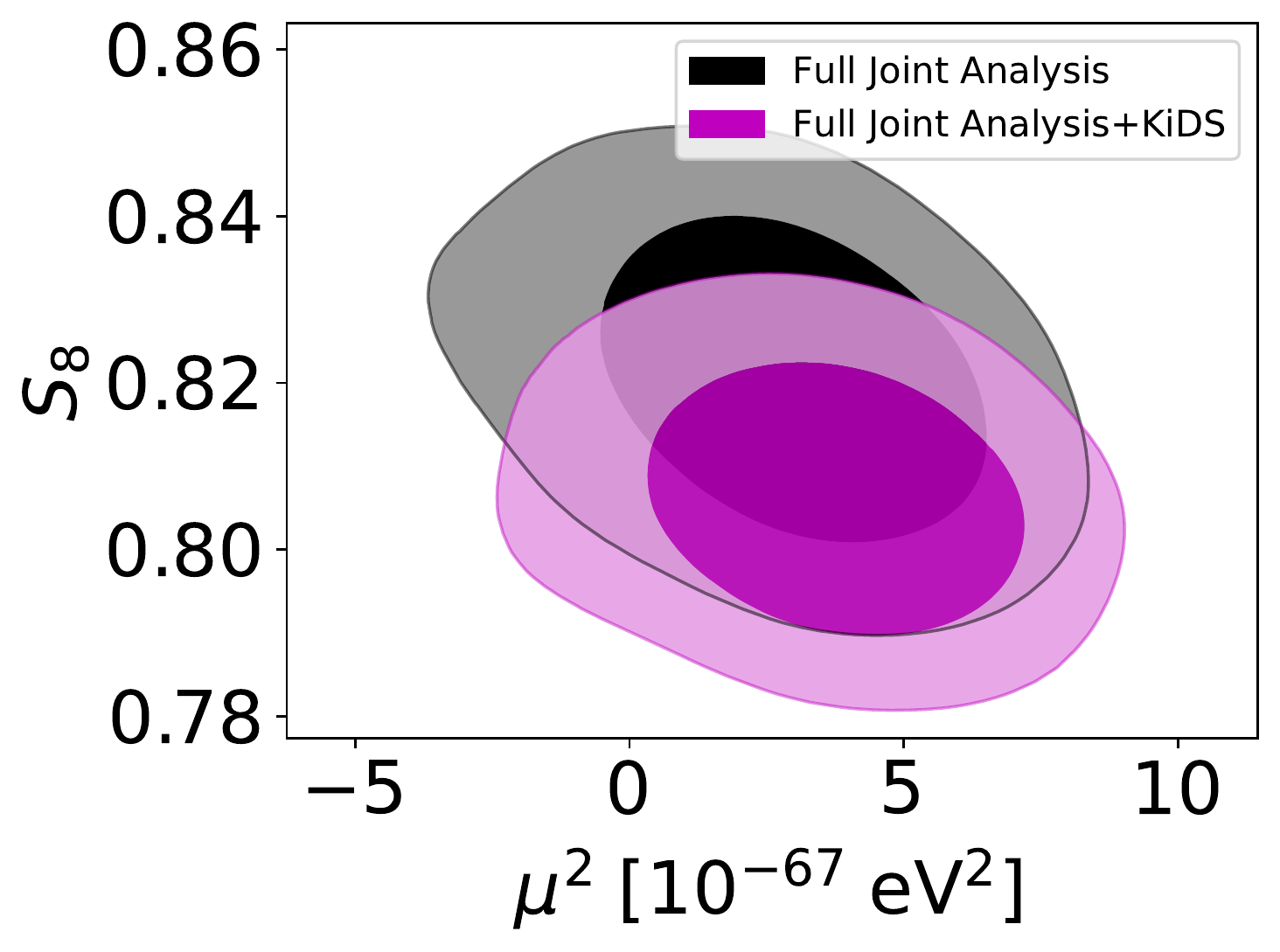}
\caption{Left panel: 68\% and 95\% CLs on the parametric space $S_8-\mu^2$ from our full joint analysis combined with a Gaussian prior on $S_8$ from DES results \cite{descollaboration2021dark}. Right panel: Same as in left panel, but the $S_8$ prior from KiDS-1000 \cite{Asgari_2021}.}
\label{S8_prior_mu}
\end{center}
\end{figure*}

Our constraints are consistent with the bound $\mu < 10^{-23}$ eV on the graviton mass set by the LIGO collaboration  \cite{theligoscientificcollaboration2020tests}.  Bounds from the solar system tests provide $\mu < 10^{-24}$ eV \cite{Will_2018}. Analyzing galactic dynamics under n-body simulations, it was shown that $\mu < 10^{-26}$ eV \cite{Brandao_2012}. Constraints from observation-derived energy condition bounds show that $\mu < 10^{-31}$ eV \cite{Alves_2018}. Recently, using 14 well-timed binary pulsars, from their intrinsic orbital decay rates, the authors in \cite{Shao_2020} found $\mu < 10^{-28}$ eV. Primordial Gravitational Waves modeled with a massive graviton will induce extra effects on the B-mode polarization of the CMB at low $l$ scales, which can place a bound on the massive graviton to $\mu < 10^{-30}$ eV \cite{Dubovsky_2010}. A more strong bound can be imposed from weak gravitational lensing observations \cite{CHOUDHURY_2004} and lunar laser ranging experiments \cite{Dvali_2003}, viz., $\mu < 10^{-32}$ eV. In Table I in \cite{de_Rham_2017}, the authors provide a compilation of several bounds on the graviton mass. Although found for a given (but realistic and stable) model describing a massive graviton, our constraint on the graviton mass is complementary and some orders of magnitude stronger than the previous ones. We refer to \cite{De_Felice_2017} for other bounds on $\mu$ in MTMG.

\section{Final Remarks}
\label{final}

Despite of the $\Lambda$CDM successes, some challenges at both the theoretical and the observational levels have placed the $\Lambda$CDM cosmology in  crossroads.  At the observational level, it faces in particular the $H_0$ and growth tensions. The growth tension is based on the fact that the observed growth of cosmological perturbations at low $z$ is weaker than the growth predicted by the standard Planck-$\Lambda$CDM parameter values.  In this work, we have investigated in detail the MTMG model in the light of different observational data sets. In particular, we have derived constraints on the MTMG using Planck-CMB data for the first time. From our full joint analysis, we have found $\mu^2/H_0^2=0.25^{{+}0.16}_{{-}0.10}$, $\mu$ being the mass of the graviton. It represents a non-null measurement on $\mu$ at 68\% CL. 

From a theoretical side, in obtaining our results, we have considered the scalar and tensor perturbative effects under a FLRW metric, while the background evolution is equivalent to $\Lambda$CDM model as described in section \ref{model}. We know that the $H_0$ value from CMB data depends on the angular scale $\theta_{*} = d_s^{*}/D_A^{*}$, where $d_s^{*}$ is the sound horizon 
at decoupling (the distance  traveled by a sound wave from the big bang to the epoch of the  CMB-baryons decoupling) and $D_A^{*}$ is the angular diameter distance at decoupling, which in turn depends on the expansion history, $H(z)$, after decoupling, controlled also by the ratio 
$\Omega_m/\Omega_{de}$ and $H_0$ mainly. In its simplest form, the MTMG scenario does not lead to changes in $H(z)$ evolution with regard to $\Lambda$CDM model. Thus, significant changes on $H_0$ that are obtained in other models in the literature, are not observed here. All changes on $H_0$ as well as on the full baseline are due to the changes on the evolution of the scalar potentials $\phi$ and $\psi$ in the Einstein's field equations. Within this framework, we have found that the current $H_0$-tension at 4.4$\sigma$ can be reduced to 3.5$\sigma$. It does not completely solve this tension, but leads us to conclude that the possibility for a non-null $\mu$ can affect the cosmological parameters estimation to CMB level too. Going beyond the simplest implementation of the MTMG model, namely changing also $H(z)$, may address this point more significantly. It may be interesting, in a future communication, to verify massive theories with change on the background evolution, which certainly should reduce the $H_0$ tension significantly. We also conclude that the well-known tension on the $S_8$ parameter can be solved in MTMG model, as shown in the previous section.
During the final stage of preparation of this work, the weak lensing and galaxy clustering measurements from the Dark Energy Survey (DES) have been updated, providing new and robust estimates on $S_8$, viz.,  $S_8 = 0.776^{+0.017}_{-0.017}$ and $S_8=0.812^{+0.012}_{-0.012}$ (DES + CMB) \cite{amon2021dark, descollaboration2021dark}, within a $\Lambda$CDM baseline. It is interesting to interpret that the MTMG scenario, without using the CMB data, predicts $S_8=0.802 \pm 0.06$, while enlarging the error bar a bit, in general, on the $S_8-\Omega_m$ plane compared to $\Lambda$CDM. Thus, comparing with DES constraint within $\Lambda$CDM baseline results, we can postulate that MTMG can also solve the $S_8$ tension between DES and CMB, which is around 2.3$\sigma$. 

We simulate an addition of the DES and Kilo-Degree Survey (KiDS-1000) data on our Full Joint Analysis by adding Gaussian priors on $S_8$ to our analysis. Figure \ref{S8_prior_mu}, on the left panel, shows the parameter space in the $S_8-\mu^2 $ plane under the prior $S_8 = 0.776 \pm 0.017$ from the DES results \cite{descollaboration2021dark}. On the right panel, we show the results assuming the prior $S_8 =0.759^{+0.024}_{-0.021}$ from the KiDS-1000 \cite{Asgari_2021}. In both cases, we note that the amplitude in $S_8$ decreases, and slightly extends the expectation of the graviton mass, but consistent with all other results developed here. Of course, an analysis within MTMG baseline using the full DES and KiDS likelihoods must confirm this prediction. We hope to provide such investigation in a future communication.

Therefore, in this work, we conclude that the MTMG is very well consistent with the CMB observations as well as to other observational ones that are sensitive to the cosmological perturbation theory. Undoubtedly, the MTMG model is a viable candidate among the modified gravity theories. \\

\begin{acknowledgments}
\noindent 

\end{acknowledgments}
J.C.N.A.\ thanks FAPESP (2013/26258-4) and CNPq (308367/2019-7) for partial financial support. The work of A.D.F.\ was supported by Japan Society for the Promotion of Science Grants-in-Aid for Scientific Research No.\ 20K03969. S.K.\ gratefully acknowledges the support from SERB-DST project No.\ EMR/2016/000258. R.C.N.\ would like to thank the agency FAPESP for financial support under the project No.\ 2018/18036-5.  Part of the numerical computation in this work was carried out at the Yukawa
Institute Computer Facility.

\bibliography{MTMG}

\end{document}